\documentclass{emulateapj}
\usepackage{amssymb, amsmath, epsfig}

\def\apj{ApJ}

\def\apjs{ApJS}
\def\aj{AJ}
\def\nat{Nature}
\def\mnras{MNRAS}

\def\aap{{A\&A}}

\begin{document}

\title{The Implications of Gunn-Peterson Troughs in the HeII Lyman-$\alpha$ Forest}
\author{Matthew McQuinn\altaffilmark{1,}\altaffilmark{2}}

\altaffiltext{1} {Harvard-Smithsonian Center for Astrophysics, 60 Garden Street, Cambridge, MA 02138, USA}
\altaffiltext{2} {Department of Astronomy, University of California, Berkeley, CA 94720, USA; mmcquinn@berkeley.edu}

\begin{abstract}
Many experts believe that the $z \sim 3$ HeII Ly$\alpha$ forest will suffer from the same saturation issues as the $z \sim 6$ HI Ly$\alpha$ forest and, therefore, will not be a sensitive probe of HeII reionization.  However, there are several factors that make HeII Ly$\alpha$ absorption more sensitive than HI Ly$\alpha$.   We show that observations of HeII Ly$\alpha$ and Ly$\beta$ Gunn-Peterson troughs can provide a relatively model-independent constraint on the volume-averaged HeII fraction of ${x}_{\rm HeII, V} \gtrsim 0.1$.  This bound derives from first using the most underdense regions in the HeII forest to constrain the local HeII fraction and, then, assuming photoionization equilibrium with the maximum allowed photoionization rate to calculate the ionization state of nearby gas.  It is possible to evade this constraint by a factor of $\sim 2$, but only if the HeII were reionized recently.   We argue that HeII Ly$\alpha$ Gunn-Peterson troughs observed in the spectra of Q0302-003 and HE2347-4342 signify the presence of $\gtrsim 10\,$ comoving Mpc patches in which ${x}_{\rm HeII, V} > 0.03$.  This is a factor of $20$ improvement over previous constraints from these spectra and $100$ times stronger than the tightest constraint on the HI volume-filling fraction from the $z>6$ HI Lyman forest.
\end{abstract}


\section{introduction}
Most experts do not believe that effective optical depths of $\tau_{\rm eff}\sim5$ measured in the $z\approx6$ HI Ly$\alpha$ forest imply that HI reionization was occurring (e.g. \citealt{becker07}).  Observations of the $z\approx3$ HeII Ly$\alpha$ forest also find $\tau_{\rm eff}\sim5$ in select regions, and again the view of experts is that these observations do not necessarily indicate that HeII reionization is happening (e.g. \citealt{fardal98, miralda00}).  However, there are several key differences that make HeII Ly$\alpha$ absorption at $z\approx 3$ a more sensitive probe of diffuse HeII than HI Ly$\alpha$ absorption at $z>6$ is of diffuse HI:  1) the abundance of helium is $\approx 14$ times smaller than that of hydrogen, 2) a photon redshifts through the HeII resonance $4$ times faster than the HI resonance, 3) the intergalactic medium (IGM) is less dense and has more voids at $z\approx 3$ than $z\approx6$,
and 4) the HI Ly$\alpha$ forest absorption is available for each HeII sightline and reveals the sightline's density structure.

Previous theoretical studies of the HeII Ly$\alpha$ forest modeled this absorption as a set of discrete absorbing clouds and/or focused on the $\tau_{\rm eff}$ statistic \citep{giroux97, fardal98, miralda00}.  This Letter employs better-suited techniques to study HeII Ly$\alpha$ absorption near saturation.  The discrete-cloud approximation works best in overdense regions.  The low-density IGM is better approximated as a less clumpy continuum of gas that has not decoupled from the Hubble flow, and it is the absorption in the least dense regions that are the most constraining with regard to HeII reionization.  Furthermore, the local measure for the HeII fraction studied here is more powerful than global measures like $\tau_{\rm eff}(z)$, for which the effect of HeII reionization is highly model-dependent.  Our measure uses HI absorption to target the HeII absorption in the most evacuated voids.  The advantages of using the HI absorption were realized in many observational studies of the HeII forest \citep{ davidsen96, hogan97, anderson99, heap00, smette02}, and we argue here that this information can be used to obtain tighter constraints on the amount of intergalactic HeII than previous analyses have found.

This study is timely because the Hubble Space Telescope (HST) reservicing mission installed the Cosmic Origins Spectrograph (COS) in May 2009.  COS will enable absorption measurements at $2.76 < z < z_{\rm QSO}$ for HeII Ly$\alpha$ and at $3.46 < z < z_{\rm QSO}$ for HeII Ly$\beta$, where $z_{\rm QSO}$ is the redshift of the quasar.  It is able to achieve higher signal-to-noise ratios and higher spectral resolutions than previous instruments on $\sim 4$ existing HeII Ly$\alpha$ sightlines, in addition to providing $\gtrsim 10$ new sightlines.  Many candidate sightlines have been identified as promising for COS spectroscopy  \citep{syphers09}.

In this Letter, we assume a flat $\Lambda$CDM cosmology with $h =0.71$, $\Omega_b = 0.046$, $\Omega_m = 0.27$, $\sigma_8 = 0.8$, and $Y_{\rm He} = 0.24$, consistent with recent measurements \citep{komatsu08}.  

\section{The HeII Lyman-series Forest}
\label{sec1}

\begin{figure}
\begin{center}
\epsfig{file=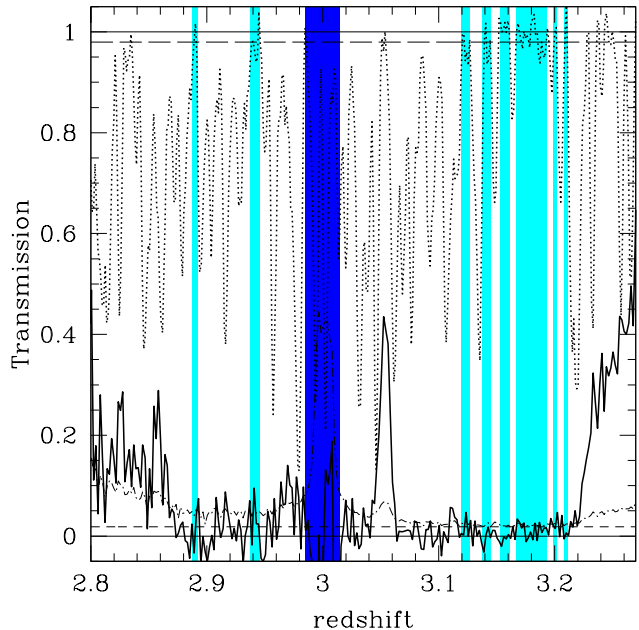, width=8.8cm}
\end{center}
\caption{HI Ly$\alpha$ forest spectrum (dotted curve) and HeII Ly$\alpha$ forest spectrum (solid curve) of Q0302-003 taken with FORS2 on the VLT and STIS on the HST, respectively \citep{worseck06}.  The dot-dashed curve is the 1$\sigma$ uncertainty on the STIS measurement.  Modulo uncertainty in the continuum level, the long-dashed curve signifies $98\%$ transmission and the short-dashed $\tau = 4$ or $\approx2\%$ transmission.  The cyan highlighted regions delineate underdense regions (as indicated by the HI data) in which the HeII Ly$\alpha$ absorption is saturated, and the dark blue region is contaminated by geocoronal Ly$\alpha$ emission. \label{fig:spectra}}
\end{figure}

In contrast to HI Ly$\alpha$ Gunn-Peterson absorption at $z \approx 6$, which saturates for neutral fractions of $\sim 10^{-5}$, the Gunn-Peterson optical depths for a photon to redshift through the HeII Ly$\alpha$, Ly$\beta$ and Ly$\gamma$ resonances are
\begin{equation}
 \tau_{\rm HeII}^{\rm GP}(\alpha, \beta, \gamma) = (3.4, 0.7, 0.2) \; \left(\frac{x_{\rm HeII}}{0.01}\right)\left( \frac{1+z}{4} \right)^{3/2} \left( \frac{\Delta_b}{0.1} \right),
 \label{eqn:GP_HEII}
\end{equation}
where $\Delta_b$ is the gas density in units of the cosmic mean and $x_{\rm HeII}$ is the local fraction of helium in HeII.\footnote{Note that damping wing absorption from HeII in dense systems or from intergalactic HeII is not significant.}  Lower redshift HeII Ly$\alpha$ absorption obscures the HeII Ly$\beta$ absorption (and both Ly$\alpha$ and Ly$\beta$ obscure the Ly$\gamma$).  If the foreground Ly$\alpha$ absorption is from $z < 2.8$ (where $\tau_{\rm eff, Ly\alpha} \lesssim 1$), this obscuration should not significantly weaken bounds that ignore it.

The focus of this Letter is on absorption in underdense regions and what this absorption implies about the ionization state of denser, neighboring regions.   Approximately 13\% of the volume has $\Delta_b < 0.2$ at $z=3$, 5\% has $\Delta_b < 0.15$ \citep{miralda00},  and a larger fraction in redshift space.  Equation~(\ref{eqn:GP_HEII}), combined with the knowledge that the least dense regions in the $z\sim 3$ IGM have $\Delta_b \sim 0.1$, suggest that the HeII Ly$\alpha$ forest is sensitive to HeII fractions in such underdensities of the order $1\%$ (and higher series lines to $\sim 10\%$).\footnote{Underdense regions will be expanding faster than the Hubble flow such that Equation~\ref{eqn:GP_HEII} will be an overestimate (by we estimate from cosmological simulations $\lesssim 50 \%$ for $\Delta_b \lesssim 0.15$).  Accounting for peculiar velocities does not change our constraints on $x_{\rm HeII}$ because peculiar velocities have the same effect on both $\tau_{\rm HeII}$ and $\tau_{\rm HI}$.}

In order to achieve such constraints, the locations of underdense regions must be known. 
The HI Ly$\alpha$ forest reveals this information.  The HI Ly$\alpha$ Gunn-Peterson optical depth is 
\begin{equation}
\tau_{\rm HI}^{\rm GP}(\alpha) \approx 0.7 \, \Delta_b^{2-0.7 (\gamma - 1)} \left(\frac{T_0}{20,000}\right)^{-0.7}  \Gamma_{12}^{-1} \left(\frac{1 +z}{4}\right)^{{9}/{2}},
\label{eqn:tauHI}
\end{equation} 
where we have assumed photoionization equilibrium with $\Gamma_{12}$ -- the HI photoionization rate in units of $10^{-12}$~s$^{-1}$ --, and a power-law temperature-density relation given by $T = T_0 \, \Delta_b^{\gamma-1}$ where $\gamma < 1.6$ \citep{gnedin98}.  The term $\Delta_b^{0.7 (\gamma - 1)}$ in equation~\ref{eqn:tauHI} arises from the temperature dependence of the recombination rate.  We use the values $\gamma = 1.3$, $T_0 = 18,000\,$K, and $\Gamma_{12} = 0.8$, consistent with observations \citep{mcdonald01b, faucher08}.  $\Gamma_{12}$ is expected to be nearly spatially invariant owing to the long mean free path (mfp) of hydrogen-ionizing photons and the large number of sources within a mfp.

The smaller the value of $\Delta_b$ that can be reliably located in the HI Ly$\alpha$ forest, the better one can constrain the HeII fraction from the HeII Lyman forest.  However, continuum fitting tends to artificially remove flux such that the flux in the lowest density pixels is set to zero, thereby preventing one from distinguishing between small values of $\Delta_b$.  \citet{faucher08b} estimated that continuum fitting on average removes $3$\% of the transmission at $z = 3$.  Therefore, a value of $\Delta_b = 0.15$, which yields $\tau_{\rm HI}^{\rm GP} \approx 0.03$ or $3\%$ absorption, is approximately the minimum density contrast over $0$ that can be discriminated at $z\sim 3$.  While a more rigorous derivation of the minimum $\Delta_b$ would be worthwhile, we employ $\Delta_b = 0.15$ for this study. 

Figure \ref{fig:spectra} shows the HI Ly$\alpha$ spectrum (dotted curve, $R  \equiv \lambda/ \Delta \lambda = 1000$) and HeII Ly$\alpha$ spectrum (thick solid curve, $R=800$) of the $z=3.29$ quasar  Q0302-003 \citep{worseck06}, one of the most-studied HeII Ly$\alpha$ sightlines.  Note that for $3.1< z < 3.2$ -- a region spanning $\approx 100$ comoving Mpc (cMpc) --  $\tau < 4$ (except perhaps near $z = 3.1$) such that there is no significant HeII Ly$\alpha$ transmission.   The short-dashed horizontal line in Figure \ref{fig:spectra} represents the transmission for $\tau = 4$.

Many of the elements with $\tau < 4$ in HeII Ly$\alpha$ absorption admit $\sim 100\%$ HI Ly$\alpha$ transmission (see the cyan highlighted regions in Figure \ref{fig:spectra}). These regions should have $\Delta_b \lesssim 0.15$ from the above discussion.  Equation (\ref{eqn:GP_HEII}) and the limit $\tau < 4$ implies that the underdense elements with $\Delta_b < 0.15$ have $x_{\rm HeII} > 0.008$.\footnote{We have checked that this bound holds if we instead use the $R = 40,000$ Keck HIRES HI Ly$\alpha$ forest spectrum of Q0302-003.  Since underdense regions tend to have several hundred km s$^{-1}$ widths, their character does not change with higher resolution data.}  Throughout this Letter, we adopt this bound on $x_{\rm HeII}$ for the most underdense, saturated elements in the HeII Ly$\alpha$ spectrum of Q0302-003, although a more detailed analysis could improve upon it.   Another well-studied HeII forest sightline, the $z=2.89$ quasar HE2347-4342 (not shown), has a Gunn-Peterson trough at $z\approx 2.85$ with no detected HeII Ly$\alpha$ transmission and provides a similar bound.

\section{HeII Photoionization}
\label{sec:photo}

\begin{figure}
\begin{center}
\epsfig{file=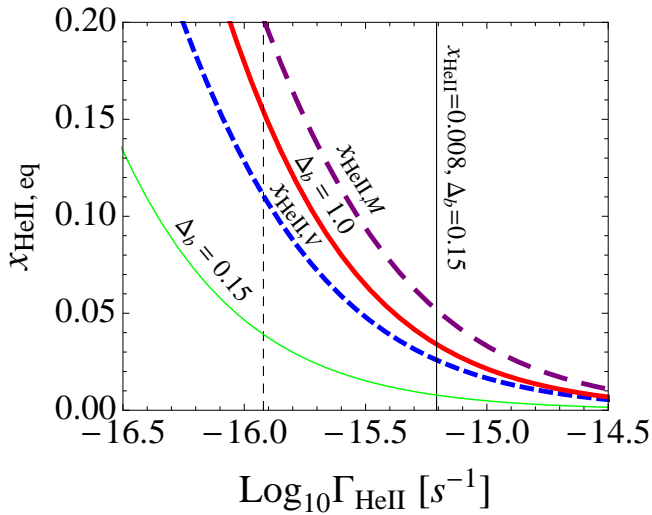, width=8.7cm}
\end{center}
\caption{HeII fraction of a patch in the $z = 3$ IGM as a function of the value of $\Gamma_{\rm HeII}$, assuming photoionization equilibrium.  The thin and thick solid curves are the HeII fraction at $\Delta_b=0.15$ and $\Delta_b=1$.  The short- and long-dashed curves are respectively the volume- and mass-weighted HeII fractions, including only elements with $\Delta_b<10$.  The intersection of the solid vertical line (at $\Gamma_{\rm HeII} \approx 6\times 10^{-16}\, {\rm s}^{-1}$) with these curves sets the lower bound on the respective HeII fraction if regions with $\Delta_b=0.15$ have $\tau > 4$ (our constraint).  The dashed vertical line is a plausible future constraint that is $5$ times stronger. 
\label{fig:HeIIfrac}}
\end{figure}

Sometime before $z \approx 2.8$, the second electron of intergalactic helium was reionized by a source of ultraviolet photons (most probably quasars), and afterward the helium was kept doubly ionized by the metagalactic radiation background.   If a gas parcel is exposed to a radiation background with HeII photoionization rate $\Gamma_{\rm HeII}$, the HeII fraction as a function of elapsed time $\Delta t$ is
\begin{equation}
  x_{\rm HeII}(t) \approx x_{\rm HeII, eq} + \left(x_{\rm HeII, 0}-  x_{\rm HeII, eq} \right) \; \exp\left(-\Delta t /t_{\rm eq}\right),
\label{eqn:xHeII}
\end{equation}
where $x_{\rm HeII, 0}$ is the HeII fraction at $\Delta t=0$, $t_{\rm eq} \equiv (\Gamma_{\rm HeII} + \alpha \, n_e)^{-1}$, 
\begin{equation}
x_{\rm HeII, eq} \equiv \frac{\alpha \, (T) \; n_e}{\Gamma_{\rm HeII} +
\alpha \, (T) \; n_e},
\label{eqn:equil}
\end{equation}
$n_e$ is the electron density, and $\alpha(T)$ is the recombination coefficient.\footnote{Note that $\Delta_b$ appears implicitly in $\alpha \, (T) \; n_e$ and $\alpha$ is the case A recombination coefficient. 
}  The $z\sim 3$ recombination time $[\alpha(T) \, n_e]^{-1}$ is roughly half the Hubble time $H(z)^{-1}$ at mean density.
Figure \ref{fig:HeIIfrac} plots $x_{\rm HeII, eq}$ as a function of $\Gamma_{\rm HeII}$ for  $\Delta_b=0.15$ and $\Delta_b=1$.  


\section{A New Method to Estimate $x_{\rm HeII}$}
\label{sec:argument}

\begin{figure}
\begin{center}
\rotatebox{-90}{\epsfig{file=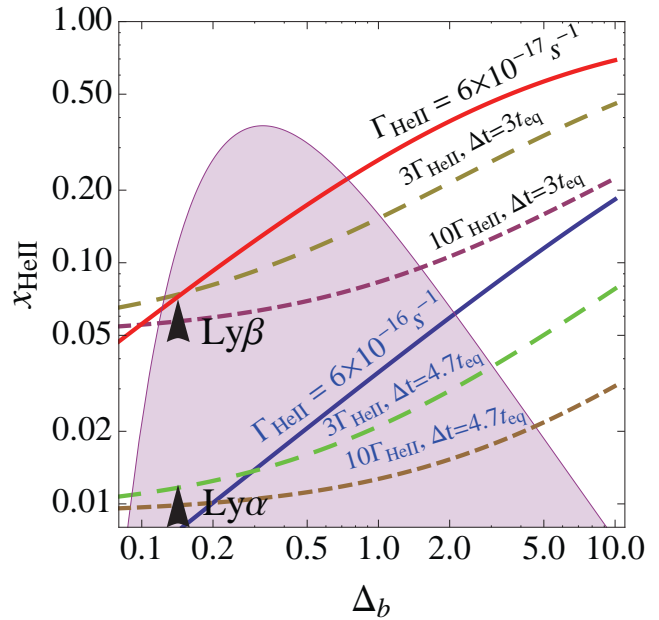, width=8.5cm}}
\end{center}
\caption{Illustration of possible scenarios in the $\Delta_b$ versus $x_{\rm HeII}$ plane at $z = 3$.  The solid curves assume equilibrium with $\Gamma_{\rm HeII} = 6\times10^{-17}~$s$^{-1}$ and $6 \times10^{-16}$ s$^{-1}$.  The long- (short-) dashed curves take $\Gamma_{\rm HeII}$ to be $3$ ($10$) times these values, but assume that equilibrium has not been reached and $x_{\rm HeII, 0} = 1$.  The arrowheads represent this Letter's constraint and an estimated future constraint, probably using Ly$\beta$.  The shaded region is $\Delta_b$ times the gas density PDF, arbitrarily normalized. \label{fig:xD}}
\end{figure}

We have shown that a measurement with percent-level precision of the HeII Ly$\alpha$ forest can be used to set a percent-level constraint on the HeII fraction in underdense regions that are opaque to HeII Ly$\alpha$ photons. In this section, we argue that such a measurement can place an even stronger limit on the local HeII fraction in a $\gtrsim 10\,$cMpc region surrounding such underdensities.  

 Our argument requires two propositions to hold:
\begin{enumerate} 
\item The ionization state of gas elements in a volume that is exposed to the same $\Gamma_{\rm HeII}(t)$ scales as $x_{\rm HeII} \propto \Delta_b^{1 - 0.7 (\gamma - 1)}$, the scaling expected for photoionization equilibrium.\footnote{This proportionality omits the $\Delta_b$ factors that appear in the denominator of Equation~\ref{eqn:equil} for notational convenience (see footnote 5).  Our calculations include this additional scaling, which increases in importance with density.}
\item A $\gtrsim 10\,$cMpc region surrounding an underdensity in the forest is exposed to approximately the same  $\Gamma_{\rm HeII}(t)$ as the underdensity.
\end{enumerate}
 
From these two propositions, one can place an interesting lower bound on $x_{\rm HeII}$ in $\gtrsim 10\,$cMpc regions.  To see this, let us take our constraint of ${x}_{\rm HeII}>0.008$ at $\Delta_b =0.15$ (Section \ref{sec1}).  Using the two propositions and equation (4) to infer the ionization state of neighboring gas elements (which provides the scaling $x_{\rm HeII} \propto \Delta_b^{1 - 0.7 (\gamma - 1)}$), we find that nearby elements with $\Delta_b = 1$ should have ${x}_{\rm HeII} > 0.034$.  See the intersection of the solid vertical line in Figure \ref{fig:HeIIfrac} with the thick solid curve.  In addition, the intersection of this vertical line with the two dashed curves in Figure \ref{fig:HeIIfrac} represents the lower limit on the volume- and mass-weighted HeII fractions in the region surrounding the opaque voids with ${x}_{\rm HeII}>0.008$ at $\Delta_b=0.15$, or ${x}_{\rm HeII, V}>0.026$ and ${x}_{\rm HeII, M}>0.052$ (excluding $\Delta_b > 10$ in these averages).  This calculation of the volume- and mass-weighted HeII fractions uses the gas density probability distribution function (PDF) measured from simulations in \citet{miralda00},\footnote{This PDF agrees well at $z\approx 3$ with more recent simulations with an updated cosmology in the range $0.1 \lesssim \Delta_b \lesssim 10$  \citep{bolton09}.}  and it uses the fact that a $\gtrsim 10\;$cMpc region is fairly representative of the IGM at $z=3$.\footnote{In linear theory in the assumed cosmology, the standard deviation of $\Delta_b$ averaged over a sphere of radius $5$ ($10$) cMpc is $0.4$ ($0.2$) at $z = 3$.}
 
 It is clear that these constraints can be improved with a more careful analysis and better data.  We expect that the constraint $x_{\rm HeII, V}>0.1$ in a $\gtrsim 10$~cMpc region is possible by utilizing smaller $\Delta_b$ (probably in a statistical manner) or by using higher Lyman-series resonances.  For example, the measurement of $\tau>4$ at $\Delta_b=0.15$ in Ly$\beta$ absorption would constrain $x_{\rm HeII, V}>0.1$ (dashed vertical line in Figure~2).\\
 
We have shown that if two propositions hold, the HeII Lyman forest can place strong constraints on $x_{\rm HeII}$ in a region.  Next, we argue that the two propositions are generally satisfied.
 
 \subsection{Proposition 1: $x_{\rm HeII} \propto \Delta_b^{1 - 0.7 (\gamma - 1)}$}
\label{ass1}
 We define HeII reionization as being over in a region when $x_{\rm HeII} \ll 1$ and $\Gamma(t)$ is much greater than $\alpha \, \bar{n}_e$, such that the region never significantly recombines.
The scaling $x_{\rm HeII} \propto \Delta_b^{1 - 0.7 (\gamma - 1)}$ will apply a few equilibrium times after HeII reionization ends in such a region even if photoionization equilibrium does not to a precision of $\delta x_{\rm HeII} < \exp(-\Delta t \langle \Gamma \rangle)$ (Equation~\ref{eqn:xHeII}), where $\langle \ldots \rangle$ represents a time average from the time HeII reionization ended, $\Delta t$.   Furthermore, once this scaling is achieved \emph{after} HeII reionization, this scaling is maintained even if photoionization equilibrium is not.

During or a few $t_{\rm eq}$ after HeII reionization this scaling may not be achieved, but it is still difficult to significantly evade the constraint on $x_{\rm HeII}$ that assumes this scaling.  To see this, let us consider a region with uniform $\Gamma_{\rm HeII}(t)$ that has an initial HeII fraction of $x_{\rm HeII, 0} \sim 1$ and is being reionized ($x_{\rm HeII}$ is decreasing).  In this case, assuming photoionization equilibrium and using underdense regions to infer the largest allowed photoionization rate $\Gamma_{\rm HeII}^{\rm max}$ could result in an overestimate of the HeII fraction if the incident $\Gamma_{\rm HeII} > \Gamma_{\rm HeII}^{\rm max}$.  This effect is illustrated by the long- and short-dashed curves in Figure \ref{fig:xD}, which are not equilibrium curves but which represent $3$ and $10$ times larger $\Gamma_{\rm HeII}$ than the corresponding solid curve (which assumes equilibrium).   However, it is unlikely that a region will have $\Gamma_{\rm HeII} \gg \Gamma_{\rm HeII}^{\rm max} $ because the timescale to reach the new equilibrium state is $ \sim \Gamma_{\rm HeII}^{-1}$, which is typically much less than $H(z)^{-1}$.  For example, the constraint $x_{\rm HeII} = 0.1$ at $\Delta_b = 1$ and $z=3$ would imply $t_{\rm eq} < {\Gamma_{\rm HeII}^{\rm max}}^{-1} \approx 0.05\, H(z)^{-1}$.  

\begin{figure}
\begin{center}
\epsfig{file=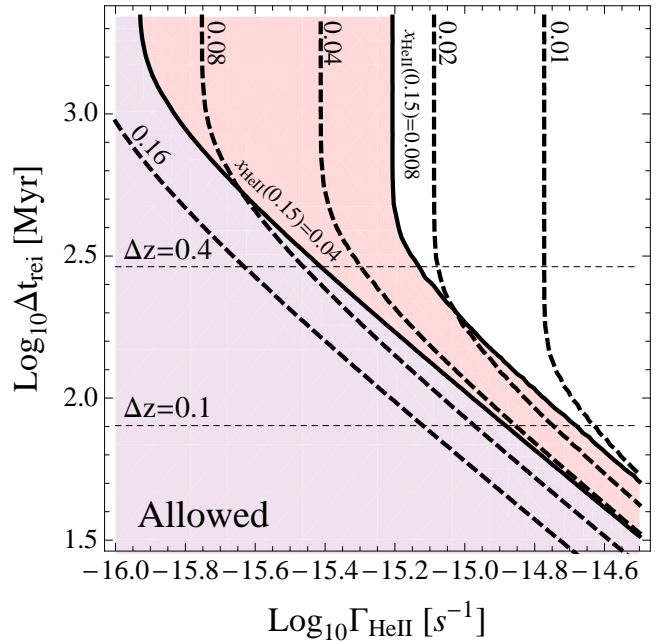, width=8.5cm}
\end{center}
\caption{Contours of fixed $x_{\rm HeII}$ in the $\Gamma_{\rm HeII}$ versus $\Delta t_{\rm rei}$ plane at $z =3$, where $\Delta t_{\rm rei}$ is the time since the beginning of HeII reionization in the region assuming constant $\Gamma_{\rm HeII}$.  The upper solid curve represents this Letter's constraint of $x_{\rm HeII} = 0.008$ at $\Delta_b= 0.15$, and the colored region left of this curve is still allowed.  The lower solid curve is the constraint that a future observation could achieve of  $x_{\rm HeII} = 0.04$ at $\Delta_b = 0.15$.  The thick dashed curves represent ${x}_{\rm HeII, V} = 0.01, ~ 0.02, ~ 0.04, ~ 0.08$, and $0.16$. \label{fig:allowed}}
\end{figure}

Figure \ref{fig:allowed} illustrates this argument for a toy example where $\Gamma_{\rm HeII}$ is taken as constant after HeII reionization begins.  Plotted are contours of fixed HeII fraction in the $\Gamma_{\rm HeII}$ versus $\Delta t_{\rm rei}$ plane, where $\Delta t_{\rm rei}$ is the time since the beginning of HeII reionization in the region.  The top thick solid curve represents our constraint $x_{\rm HeII} = 0.008$ at $\Delta_b= 0.15$.  Only the colored area left of this curve is allowed.  The dashed curves represent ${x}_{\rm HeII, V} = 0.01, ~ 0.02, ~ 0.04, ~ 0.08$, and $0.16$.  This illustrates that in order for ${x}_{\rm HeII, V}$ to be significantly less than the limit we set assuming photoionization equilibrium (or ${x}_{\rm HeII, V} = 0.026$; Section 4), the HeII needs to have been recently reionized.  For example, to evade this limit such that ${x}_{\rm HeII, V} = 0.015$, the HeII in the region must have been reionized over the last $100$ Myr ($\Delta z \approx 0.1$).

 \subsection{Proposition 2: $\Gamma_{\rm HeII}$ fluctuates on $\gtrsim 10$~cMpc}
\label{ass2}
This proposition holds because the mfp of a HeII-ionizing photon, which sets the correlation length for $\Gamma_{\rm HeII}$, should be $\gtrsim 10 \,$cMpc.  The mfp for a HeII Lyman-limit photon to be absorbed in diffuse gas at $z=3$ is $\approx 0.8 \, {x}_{\rm HeII, V}^{-1}\,$cMpc (assuming a homogeneous IGM), which is $\gtrsim 10\,$cMpc for the value of ${x}_{\rm HeII, V}$ that saturates the previous bounds we have quoted, $x_{\rm HeII, V} < 0.1$.  The spectrally-averaged mfp, weighting by contribution to $\Gamma_{\rm HeII}$, is $3$ and $7$ times larger than the Lyman-limit mfp for photon spectral indexes of $2.5$ (expected in the case of no absorption) and $1.5$, respectively.  In addition, estimates for the mfp of HeII-ionizing photons to be absorbed in dense systems {\`a} la \citet{haardt96} are similar at the $\Gamma_{\rm HeII}$ that yield $x_{\rm HeII, eq, V} < 0.1$ (Appendix A in \citealt{mcquinn09} plus followup work).  In agreement with this proposition, the HeII reionization simulations of \citet{mcquinn09} find $\Gamma_{\rm HeII}$-fluctuations on $\gtrsim 10~$cMpc scales (even in HeII regions; see their Figure~5).  

The previous paragraph argued that $\Gamma_{\rm HeII}$ fluctuates as $\gtrsim 10\,$cMpc when our bound on $x_{\rm HeII, V}$ is saturated.  However, $x_{\rm HeII, V}$ could be larger than our lower bound immediately around the underdensity such that the fluctuation scale is $< 10\,$cMpc.  One could then design a situation in which the average $\bar{x}_{\rm HeII}$ in a $10\,$cMpc region around an underdensity is less than our bound.  However, the situation would have to be very contrived in order to significantly evade our bound, especially since current spectra show saturated voids throughout their Gunn-Peterson troughs.  

In apparent contradiction to this proposition, \citet{shull04} detected large fluctuations in $\Gamma_{\rm HeII}$ on $\sim 1\,$cMpc scales in the spectra of HE2347-4342, with the least dense pixels yielding smaller $\Gamma_{\rm HeII}$.  However, \citet{bolton05} replicated the \citet{shull04} analysis on mock data with the same noise properties, and they showed that the observed fluctuations could be explained by Poisson fluctuations in the distribution of quasars and by a $30\,$cMpc mfp.   It appears the reason that a $\sim1\,$cMpc correlation length is inferred in \citet{bolton05} owes to noise in the HeII forest spectrum, which often results in a large overestimate for $\Gamma_{\rm HeII}$ in pixels with $\tau>0.05$.  In fact, \citet{fechner07} demonstrated the presence of this bias and, when it was corrected for, estimated a fluctuation scale for $\Gamma_{\rm HeII}$ of $\geq 8-24\,$cMpc from HE2347-4342.

Related to Proposition 2 is the misconception that $\Gamma_{\rm HeII}$ should correlate strongly with the density of an absorber.  The rays from a quasar will traverse many (essentially uncorrelated) absorption systems prior to traveling  $\sim 10\,$cMpc to a typical absorber, making it implausible that the typical overdense element will be exposed to a larger $\Gamma_{\rm HeII}$ from these sources than an underdense one.   This argument ignores that overdense absorbers are more likely to exist closer to a quasar where there is more radiation.  However, the correlation between the locations of $\sim L_*$ quasars ($\sim 1$ per $30^3$ cMpc$^3$) and the density of a typical absorber is negligible \citep{mcquinn09}.  Even if $\sim L_*$ quasars were not the primary source for $\Gamma_{\rm HeII}$ [contrary to what the quasar luminosity function and the amplitude of $\Gamma_{\rm HeII}$-fluctuations in the HeII forest suggest \citep{mcquinn09, bolton05}], it would be difficult to make $\Gamma_{\rm HeII}$ correlate strongly with the density of absorbers:  as the number density of the sources increases, the fluctuations in $\Gamma_{\rm HeII}$ should decrease.


\section{Conclusions}
\label{sec:taueff} \label{conclusions}

This Letter showed that current HeII Ly$\alpha$ forest data imply that ${x}_{\rm HeII, V}>0.03$ in regions surrounding opaque voids, while ${x}_{\rm HeII, V}\gtrsim0.1$ should be possible with future HeII Lyman forest data.  These limits derive from assuming photoionization equilibrium and that the maximum allowed $\Gamma_{\rm HeII}$ in an underdense gas element is also the maximum in a surrounding $\gtrsim 10\,$cMpc region -- the expected fluctuation scale for $\Gamma_{\rm HeII}$.  They may be weakened somewhat, but only if the region in question is being reionized.  However, since HeII reionization is the epoch when a non-negligible HeII fraction exists in the low-density IGM, by definition it would be occurring if ${x}_{\rm HeII, V}\gtrsim0.1$.  Therefore, the HeII forest may be able to address whether HeII reionization is occurring at $z\approx3$.

Our constraint of $x_{\rm HeII, V}>0.03$ is a significant improvement over previous efforts, which set the bound $x_{\rm HeII, V}>1.3\times10^{-3}$ in the regions of highest opacity \citep{heap00}.  Our constraint on ${x}_{\rm HeII, V}$ in select regions is $100$ times stronger than the tightest constraint on the volume-averaged HI fraction from the $z>6$ HI Ly$\alpha$--$\gamma$ forests of $\bar{x}_{\rm HI, V} \gtrsim 3\times10^{-4}$ \citep{fan06}. It can be improved by better observations, by including redshift-space effects, by measuring $x_{\rm HeII}$ in pixels with $\Delta_b<0.15$, or by utilizing absorption information from higher Lyman-series resonances.\\


We would especially like to thank Gabor Worseck for providing the STIS and FORS2 spectra and Michael Rauch for the HIRES.  We also thank Jamie Bolton, Claude-Andr{\'e} Faucher-Gigu{\`e}re, Adam Lidz, and Gabor Worseck for useful discussions.

\bibliographystyle{apj}

\end{document}